\documentclass[prb,preprint,superscriptaddress,preprintnumbers,amssymb]{revtex4}
\usepackage[english]{babel}
\usepackage[letterpaper,top=2cm,bottom=2cm,left=3cm,right=3cm,marginparwidth=1.75cm]{geometry}

\usepackage{soul}
\usepackage{comment}
\usepackage{orcidlink}

\usepackage{amsmath}
\usepackage{bm}
\usepackage{siunitx}
\usepackage{nicefrac}

\usepackage{graphicx}
\usepackage{xcolor}
\usepackage{xspace}
\usepackage{float}

\newcommand{\ket}[1]{| {#1} \rangle} 
\DeclareMathAlphabet\mathbfcal{OMS}{cmsy}{b}{n}

\usepackage{doi}
\usepackage{hyperref}
\newcommand{\SAVBA}{\affiliation{Institute of Informatics, Slovak Academy of Sciences, 84507 Bratislava, Slovakia}}
\newcommand{\FFBG}{\affiliation {Faculty of Physics, University of Belgrade, 11001 Belgrade, Serbia}}
\newcommand{\vinca}{\affiliation{Vin\v{c}a Institute of Nuclear Sciences - 
National Institute of the Republic of Serbia, University of Belgrade, P. O. Box 522, RS-11001 Belgrade, Serbia}}
\newcommand{\UPJS}{\affiliation{Institute of Physics, Pavol Jozef \v{S}af\'{a}rik University in Ko\v{s}ice, 04001 Ko\v{s}ice, Slovakia}}
\newcommand{\SAVKE}{\affiliation{Institute of Experimental Physics, Slovak Academy of Sciences, 04001 Ko\v{s}ice, Slovakia}}
\newcommand{\cnrspin}{\affiliation{Consiglio Nazionale delle Ricerche CNR-SPIN, c/o Universit\'{a} degli Studi "G. D’Annunzio", 66100 Chieti, Italy}}

\begin{document}
\title{Interplay of altermagnetism and weak ferromagnetism \\ in two-dimensional RuF$_4$}

\author{Marko Milivojevi\'c \orcidlink{0000-0002-9583-3640}} \SAVBA \FFBG
\author{Marko Orozovi\'c \orcidlink{0000-0002-3018-5830}} \vinca 
\author{Silvia Picozzi} \cnrspin
\author{Martin Gmitra \orcidlink{0000-0003-1118-3028}} \UPJS \SAVKE
\author{Srdjan Stavri\'c \orcidlink{0000-0003-2097-0955}}\email{stavric@vin.bg.ac.rs} \vinca \cnrspin 

\date{\today}

\begin{abstract}
Gaining growing attention in spintronics is a class of magnets displaying zero net magnetization and spin-split electronic bands called altermagnets. Here, by combining density functional theory and symmetry analysis, we show that RuF$_4$ monolayer is a two-dimensional $d$-wave altermagnet. Spin-orbit coupling leads to pronounced spin splitting of the electronic bands at the $\Gamma$ point by $\sim 100 \, {\rm meV}$ and turns the RuF$_4$ into a weak ferromagnet due to non trivial spin-momentum locking that cants the Ru magnetic moments. The net magnetic moment scales linearly with the spin-orbit coupling strength. 
Using group theory we derive an effective spin Hamiltonian capturing the spin-splitting and spin-momentum locking of the electronic bands. Disentanglement of the altermagnetic and spin-orbit coupling induced spin splitting uncovers to which extent the altermagnetic properties are affected by the spin-orbit coupling. Our results move the spotlight to the non trivial spin-momentum locking and weak ferromagnetism in the two-dimensional altermagnets relevant for novel venues in this emerging field of material science research.

\end{abstract}
\maketitle

\section{Introduction}

\noindent
Recent discovery of non-relativistic spin splitting in electronic bands of materials with compensated collinear magnetic order opened new venue in antiferromagnetic spintronics\cite{Zutic2004,Jungwirth2016Mar,Smejkal2018Mar}. These intriguing materials constitute a separate magnetic phase, besides ferromagnets (FM) and antiferromagnets (AF), dubbed the name altermagnets (AM) \cite{Smejkal2022Sep,Smejkal2022Dec} or, alternatively, antiferromagnets with non-interconvertible spin-structure motif pair\cite{Yuan2020Jul,Yuan2021Jan, Yuan2023Aug_AdvMater}. 
Altermagnets display a profusion of diverse physical phenomena, including anomalous Hall effect~\cite{Tschirner2023}, piezoelectricity~\cite{Guo2023}, and chiral magnons\cite{Smejkal2023Dec}; they can be used for efficient spin-to-charge conversion~\cite{Bai2023}, to generate orientation-dependent spin-pumping currents\cite{Sun2023Oct} and to proximitize superconductor materials~\cite{Ouassou2023,Vakhtel2023,Sun2023}, just to scratch the manifold of their possible applications.

Many altermagnetic compounds have been classified so far\cite{Guo2023Mar,Biniskos2023}, including MnTe~\cite{Mazin2023,Faria2023}, MnF$_2$\cite{Yuan2020Jul}, RuO$_2$ \cite{Smejkal2022Sep,Feng2022Nov,Tschirner2023} and CrSb \cite{Reimers2023Oct}. In MnTe the spin splitting of $380 \, {\rm meV}$ is experimentally determined by angle-resolved photoemission spectroscopy (ARPES) measurements, thus confirming its altermagnetic status previously predicted by density functional theory (DFT) calculations. For possible spin-transport applications, apart from the magnitude of the splitting energy, the positioning of the spin-split bands is highly important. In this respect CrSb may be a promising candidate for producing spin-polarized currents as a giant spin splitting of $\sim 600 \, {\rm meV}$ occurs just below the Fermi energy\cite{Reimers2023Oct}. 

Regarding their dimensionality, all the previously mentioned altermagnets are three-dimensional (3D) compounds, and as of now, reports on two-dimensional (2D) altermagnets are limited.
In the work of Brekke \textit{et al}~\cite{Brekke2023Dec} it is proposed that magnon-mediated superconductivity in a 2D altermagnet could result with a superconductor with substantially enhanced critical temperature. However, this is accomplished using a microscopic model, and no specific material has been suggested to test this result. Very recently, monolayer Cr$_2$Te$_2$O is predicted as a promising platform for practical realization of the spin Seebeck and spin Nernst effects of magnons\cite{Cui2023Nov}. Further, it is shown that monolayer MnPS$_3$, which is a conventional antiferromagnet, can be converted into altermagnet if one of its S layers is replaced by Se\cite{Mazin2023Sep}. However, in thus obtained MnP(S,Se)$_3$ Janus monolayer, the spin splitting of the relevant top valence band is modest, not exceeding $\approx 10 \, {\rm meV}$ in the best case. Considering this, the endeavor to fabricate the Janus monolayer might be deemed excessively demanding in comparison to the anticipated benefits. Hence, for prospective applications in miniature and highly efficient spintronic devices, it would be beneficial to have 2D materials that inherently exhibit altermagnetic properties.

There is yet an important peculiarity concerning 2D magnets in general. Namely, in the absence of an external magnetic field the long-range order of spins on a 2D lattice can sustain above $T = 0 \, {\rm K}$ temperature only if the system possesses the magnetic anisotropy energy (MAE)~\cite{Mermin1966Nov}. MAE is crucial here as it gaps the magnon spectra. Otherwise, the low-energy magnon excitations would destroy the long-range order at arbitrary non-zero temperature. In this respect, altermagnets are not any different from ferromagnets and antiferromagnets. MAE, in turn, requires the presence of a sizable spin-orbit coupling (SOC) in the system. Therefore, for a realistic description of a 2D altermagnet the SOC must be accounted for. This raises a question -- to which extent its non-relativistic properties, e.g. the hallmark spin splitting of bands, are affected by SOC?

Here, we show that in the non-relativistic limit monolayer RuF$_4$ is a 2D altermagnet and we describe the spin splitting of band structure and spin textures of the altermagnetic phase. Subsequently, we analyze the influence of SOC on the altermagnetic properties. The structure of monolayer RuF$_4$, its crystallographic and magnetic symmetries, are described in Subsection~\ref{ssec_symmetry_analysis}. The altermagnetic phase is analyzed in Subsection~\ref{ssec.nonrel_description}. Weak ferromagnetism and SOC-induced spin splitting of bands are exposed in Subsections~\ref{ssec.weak_FM} and \ref{ssec.band_splitting}. Finally, we summarize the paper in Section~\ref{sec.conclusion} with an outlook on the link between altermagnetism and weak ferromagnetism.  

\section{Results}

\subsection{The structure and crystal symmetries of RuF$_4$ monolayer}
\label{ssec_symmetry_analysis}
\begin{figure}[h!]
\centering
\includegraphics[width=1.0\linewidth]{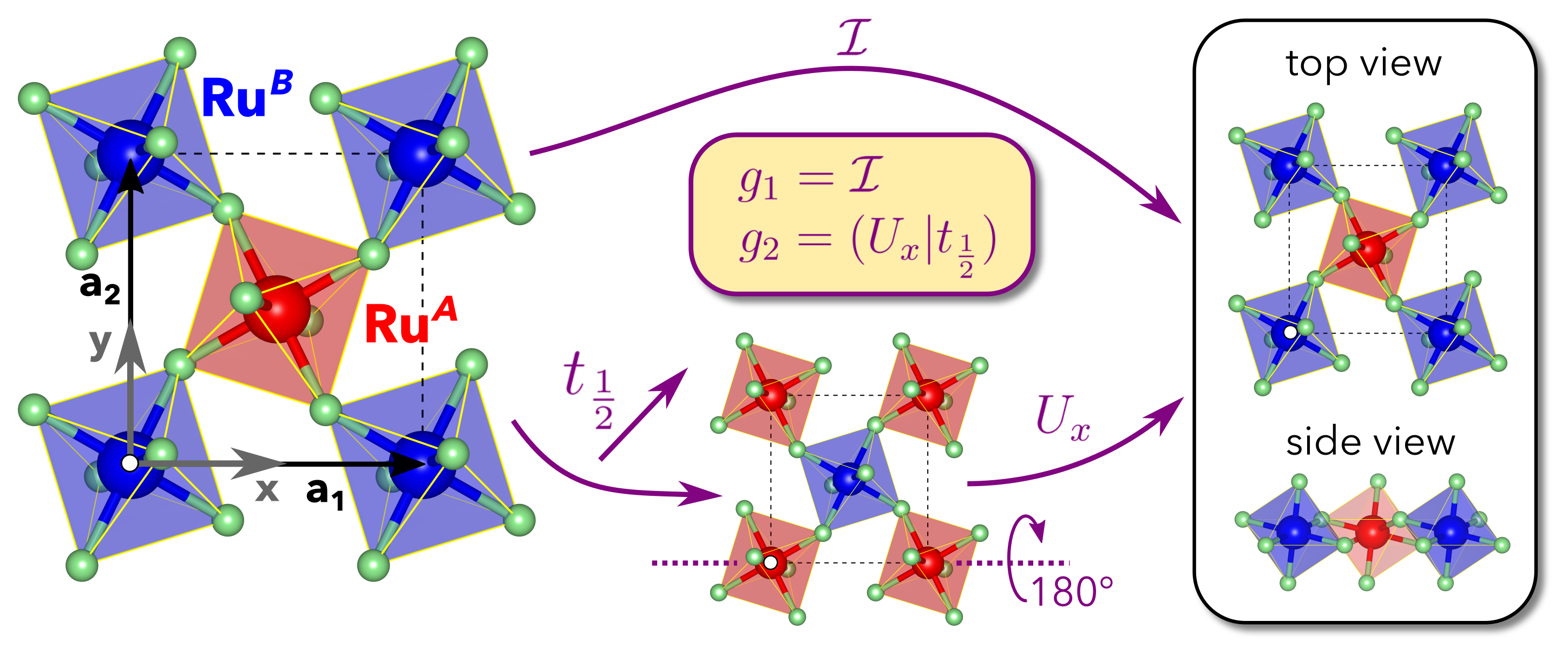}
\caption{
Structure of monolayer RuF$_4$ and a schematic depiction of transformations $g_1=\mathcal{I}$ and $g_2=(U_x|t_{\frac{1}{2}})$ that are generators of its crystallographic group. Distinct Ru clusters are depicted by red (Ru$^A$) and blue (Ru$^B$) octahedra. 
}
\label{fig:structure} 
\end{figure}

\noindent
Monolayer RuF$_4$ is composed of ruthenium atoms sitting in centers of fluorine octahedra. It is a van der Waals material that can be exfoliated from the bulk compound which was experimentally synthesized in monoclinic $P2_1/c$ crystallographic space group (group No. 14) \cite{Casteel1992Jul}.  
We will refer to an Ru atom with its surrounding F ligands as \textit{the Ru cluster}. In each Ru cluster two out of six F atoms belong only to that cluster, whereas four of them are shared with four nearest neighbors. Crystal lattice of monolayer RuF$_4$ has two sublattices with symmetrically distinct Ru clusters, Ru$^A$ and Ru$^B$, that are arranged as depicted in Fig.~\ref{fig:structure}.  

Two perpendicular lattice vectors that define the RuF$_4$ plane, ${\bf a}_1$ and ${\bf a}_2$, have different norm, $a_1 = 5.131 \, {\rm \AA}$ and $a_2 = 5.477 \, {\rm \AA}$, as obtained from our DFT calculations. Therefore, we will refer to the two directions defined by these vectors as \textit{the shorter} and \textit{the longer} in-plane directions. 
The structure of monolayer RuF$_4$ belongs to the $p2_1/b11$ layer group (layer group No. 17)\cite{Litvin1991} which has two generators: the space inversion $g_1=\mathcal{I}$ and the non-symmorphic group element $g_2=(U_x|t_{\frac{1}{2}})$ that is the combination of two operations each transforming Ru$^A$ into Ru$^B$ and vice versa: a rotation $U_x$ by $180^\circ$ around the shorter in-plane direction ${\bf a}_1$ and a fractional lattice translation $t_{\frac{1}{2}} = \frac{1}{2}({\bf a}_1 + {\bf a}_2)$. Therefore, the group symmetry consists of four elements, ${\bf G} = \{e,g_1,g_2,g_1g_2\}$, where $e$ is the identity element and $g_1g_2$ represents a joint action of the group generators. Importantly, the crystal structure is \textit{centrosymmetric}, but the Ru-Ru bond does not have the point of inversion, meaning that the Dzyaloshinskii-Moriya interaction can occur between the Ru magnetic moments\cite{Moriya1960Oct}. We will discuss this in detail in the remainder of the manuscript. 

\subsection{Non-relativistic description of magnetic properties: 2D altermagnet}
\label{ssec.nonrel_description}
\noindent
RuF$_4$ has a compensated net magnetic moment due to the opposite magnetic moments on the two Ru sublattices. Octahedral crystal field of F$^-$ ions acts on Ru$^{4+}$ ion, leading to the $t_{2g} - e_g$ splitting of its $4d$ orbitals and leaving two spin-unpaired electrons in the $t_{2g}$ manifold. Therefore, from the crystal field theory, magnetic moments of $2 \, \mu_{\rm B}$ are expected on Ru atoms. Our DFT calculations give $1.464 \, \mu_{\rm B}$, in agreement with the value reported in Ref.~\cite{Wang2022Aug}. The discrepancy between the expected and the DFT-calculated magnetic moments is due to the hybridization between the Ru $4d$ and the F $2p$ orbitals. Concretely, the F atoms that belong to a single Ru cluster exhibit moderate magnetic moments of 0.214 $\mu_{\rm B}$, induced by the nearest Ru magnetic moment, that are coupled ferromagnetically to that moment. On the other hand, the F atoms shared between two clusters are bonded to the two opposite-spin Ru atoms and are thus non-magnetic. In summary, the magnitude of the magnetic moment on each Ru cluster is $|{\bf m}| = |{\bf m}({\rm Ru})+2{\bf m}({\rm F})| = 1.9 \, \mu_{\rm B}$. 

In the absence of an external magnetic field and without SOC, the direction of magnetic moments and thus of the N\'{e}el vector $({\bf N})$ is arbitrary. To perform symmetry analysis, we assume that ${\bf N}$ points along the longer in-plane lattice direction ${\bf a}_2$, which coincides with the $y$-axis in our choice of the coordinate frame (see Fig.~\ref{fig:structure}). Space inversion $g_1$ has a trivial effect on magnetic moments as they are pseudovectors, whereas $g_2$ leaves the magnetic order invariant because the rotation $U_x$ and the fractional lattice translation $t_{\frac{1}{2}}$ each reverse the magnetic order. Therefore, all the symmetry operations from the $p2_1/b11$ layer group leave the magnetic order unaltered and are thus the symmetries of the magnetic system as well, meaning that the magnetic space group is equal to its parent crystallographic group (type-I). Given that (1) the magnetic space group is of type-I\cite{BC91} and (2) the combination $\theta \mathcal{I}$ of time reversal $\theta$ and space inversion symmetry $\mathcal{I}$ is broken due to the magnetic order, the \textit{non-relativistic spin splitting} in monolayer RuF$_4$ is allowed \cite{Yuan2021Jan, Guo2023Mar}. 

\begin{figure}[h!]
\centering
\includegraphics[width=0.8\linewidth]{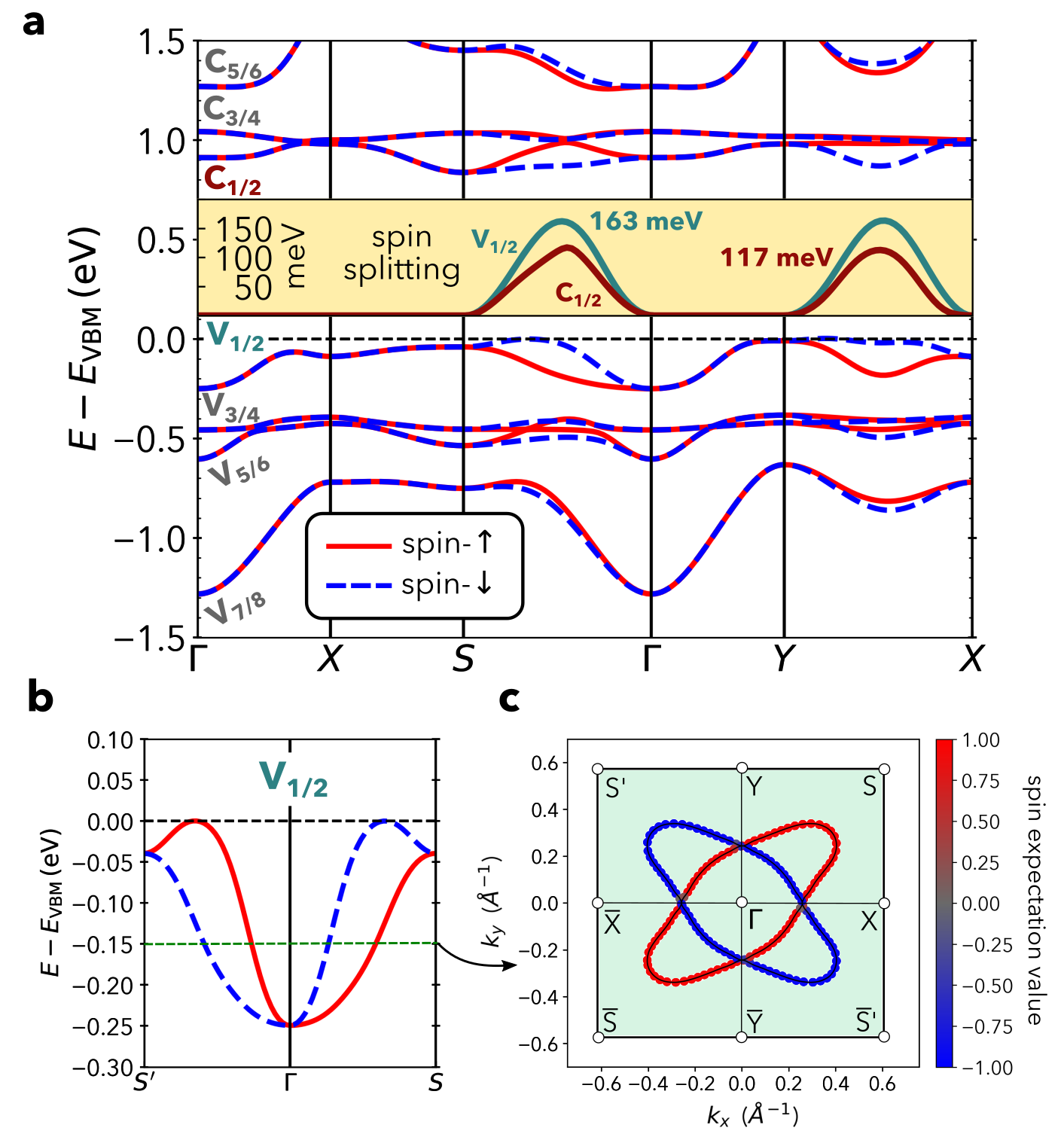}
\caption{\label{fig:bands-AM} Non-relativistic RuF$_4$ band structure calculated with spin-polarized DFT calculations. In~(a), the $\Gamma$XS$\Gamma$YX path is assumed and the spin splitting energies of the top valence (V$_{1/2}$) and bottom conduction (C$_{1/2}$) bands are given in the inset (yellow rectangle). In (b) the top valence band V$_{1/2}$ is illustrated once again along the S$' \Gamma$S path, and the $d$-wave nature of its spin-momentum locking in the full Brillouin zone (green rectangle) at $-0.15 \, {\rm eV}$ below the Fermi energy is illustrated in (c).
}
\end{figure}

In Fig.~\ref{fig:bands-AM}a we plot the band structure of monolayer RuF$_4$ calculated with spin-polarized DFT calculations. The system is a semiconductor with an indirect band gap of $0.83 \, {\rm eV}$ between the valence band maximum, located on the XY path, and the conduction band minimum located at S. 
Along the chosen $k$-path in the Brillouin zone (BZ), the spin-up and spin-down bands are degenerate along the $\Gamma$X, XS, and $\Gamma$Y lines, whereas a sizable spin splitting $\Delta E_n ({\bf k}) = E^{\uparrow}_n({\bf k})-E^{\downarrow}_n(\bf k)$ ($n$ is the band index) is observable along the  $\Gamma$S and XY lines. 
Notably, of all the bands in the energy window from $-1.5 \, {\rm eV}$ below to $1.5  \, {\rm eV}$ above $E_{\rm F}$, the spin splitting is the largest for the top valence (V$_{1/2}$ shown in Fig.~\ref{fig:bands-AM}b) and the bottom conduction bands (C$_{1/2}$), reaching $163 \, {\rm meV}$ and $117 \, {\rm meV}$, respectively (yellow rectangle in Fig.~\ref{fig:bands-AM}a). The spin-momentum locking of the bands V$_{1/2}$ presented in Fig.~\ref{fig:bands-AM}c clearly shows that monolayer RuF$_4$ is a $d$-wave altermagnet\cite{Smejkal2022Sep}. 

To model the spin splitting of these bands around the $\Gamma$ point we employ a simple time-reversal breaking Hamiltonian, compatible with the system's symmetry,
\begin{equation}
\label{SOCmodel}
\mathcal{H}_{\rm ss}=\frac{1}{2}\alpha k_x k_y \sigma_y . 
\end{equation}
Here $\sigma_y = \pm 1$ and $\alpha$ is the band-dependent parameter determined by fitting the model to the DFT bands. For the V$_{1/2}$ and C$_{1/2}$, up to the 25\% of the $\Gamma$X  line, the spin splitting in an arbitrary $k$-direction can be described using $\alpha_{V}=3450\,{\rm meV\, \AA^2}$ and $\alpha_{C}=2368\,{\rm meV\, \AA^2}$.

As DFT calculations pointed out, the spin degeneracy of bands is not lifted along the $\Gamma$X and $\Gamma$Y lines. This can be explained by analyzing how the group elements connect different points of the energy surface $E_n(k_x,k_y,s)$. Keeping the assumption that magnetic moments are in the $y$-direction, the action of group elements reflects on the electronic bands as
\begin{eqnarray}
g_1\,E_n(k_x,k_y,s)&=&E_n(-k_x,-k_y,s),\nonumber\\
g_2\,E_n(k_x,k_y,s)&=&E_n(k_x,-k_y,-s), \nonumber\\
g_1g_2\,E_n(k_x,k_y,s)&=&E_n(-k_x,k_y,-s).
\label{eq.g_action_on_energy}
\end{eqnarray}
Based on the equations (\ref{eq.g_action_on_energy}), one concludes that along $\Gamma$X, where ${\bf k}=(k_x,0)$,  and $\Gamma$Y, where ${\bf k}=(0,k_y)$, the relations
\begin{eqnarray}
g_2\,E_n(k_x,0,s)=E_n(k_x,0,-s),\nonumber\\
g_1g_2\,E_n(0,k_y,s)=E_n(0,k_y,-s)
\label{eq.XGY}
\end{eqnarray}
express the spin degeneracy. On the other hand, these constraints are not present along an arbitrary $k$-path and the spin splitting of bands is not forbidden. This is exactly the case of the $\Gamma$S and XY lines, where we observe a noticeable non-relativistic spin splitting, which is the hallmark of the altermagnetic phase. 

\subsection{SOC turning altermagnet into a weak ferromagnet}
\label{ssec.weak_FM}
\noindent

Once the SOC is included, the spin space couples to the real space and the N\'{e}el vector ${\bf N}$ obtains a defined crystallographic direction. Our DFT calculations reveal that the shorter in-plane lattice direction (the $y$-axis) is the preferential direction for ${\bf N}$, with MAE separating it from the longer in-plane direction ($x$-axis) and from the out-of-plane direction ($z$-axis) by $1.12 \, {\rm meV}/{\rm Ru}$ and $7.04 \, {\rm meV}/{\rm Ru}$, respectively. 

\begin{figure}[h]
\centering
\includegraphics[width=0.8\linewidth]{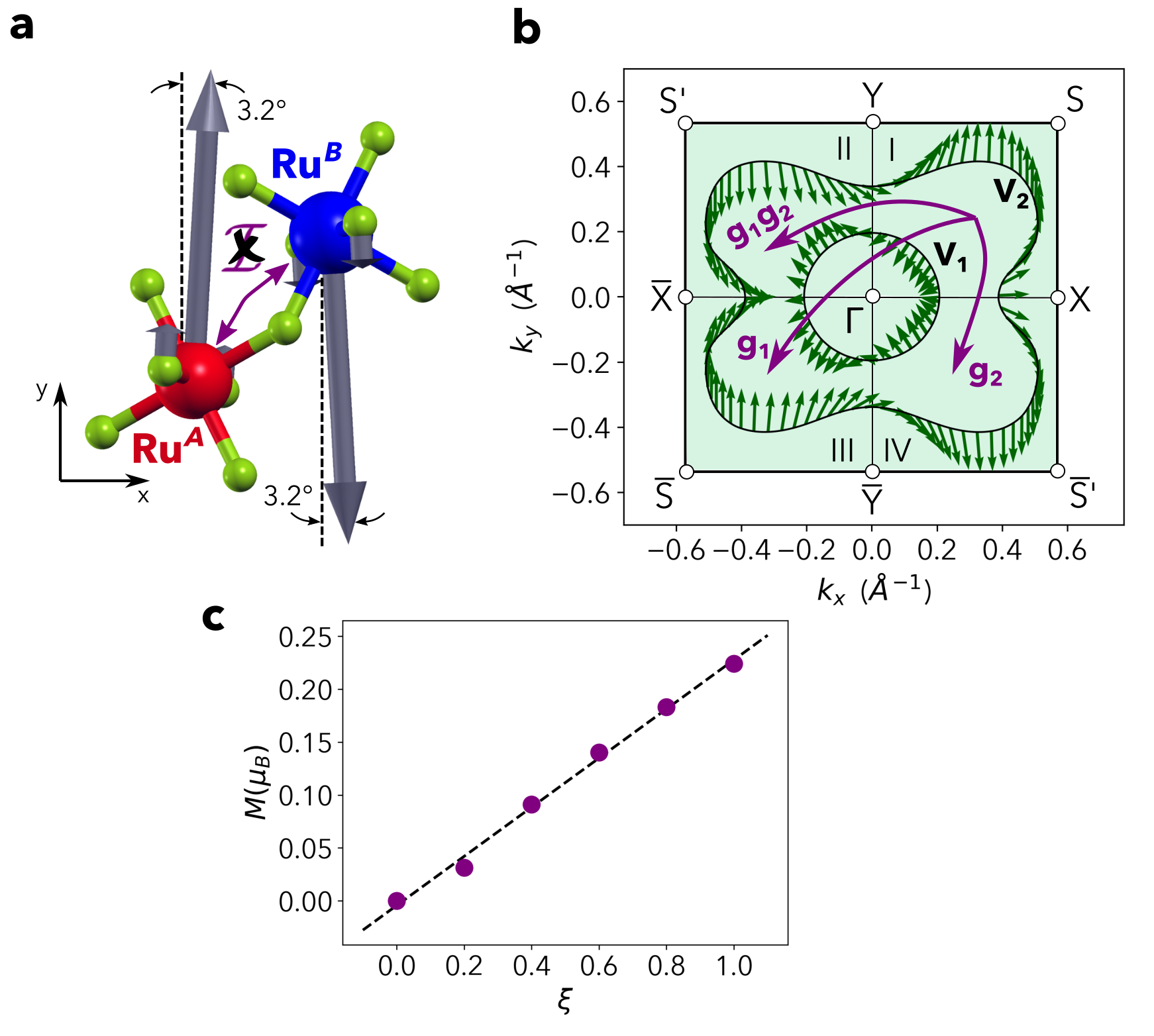}
\caption{\label{fig:canting} 
(a) SOC-induced canting of magnetic moments on two sublattices of RuF$_4$. 
(b) Spin textures of the top valence bands V$_1$ and V$_2$ at energy -0.15 eV below the Fermi level. (c) Dependence of induced net magnetic moment on the SOC strength.}
\end{figure}

Due to SOC,  the magnetic moments are getting canted from the $y$-axis towards the $x$-axis by $3.2^\circ$, 
as depicted in Fig.~\ref{fig:canting}a. Canting lowers the total energy only slightly ($0.08 \, {\rm meV}/{\rm Ru}$), but it induces a net magnetic moment of $|{\bf M}| = |{\bf m}_A + {\bf m}_B| = 0.22 \, \mu_{\rm B}$ which points in the $+x$ direction. Therefore, upon the inclusion of SOC, the monolayer RuF$_4$ changes its magnetic phase from the altermagnetic to the \textit{weak ferromagnetic} (WF) phase. 

The direction of the induced magnetic moment in RuF$_4$ can be inferred from the group theory analysis. In general, the magnetic moment ${\bf M}=(M_x,M_y,M_z)$ of any system must stay invariant under the action of all the group symmetry elements. Since the inversion $g_1$ has a trivial action on the pseudovector ${\bf M}$ through the $\mathcal{D}$ representation~\footnote{$\mathcal{D}(g_1)=\Big(\begin{smallmatrix}
1 & 0 & 0\\
0 & 1 & 0\\
0 & 0 & 1
\end{smallmatrix}\Big)$, $\mathcal{D}(g_2)=\Big(\begin{smallmatrix}
1 & 0 & 0\\
0 & -1 & 0\\
0 & 0 & -1
\end{smallmatrix}\Big)$, $\mathcal{D}(g_1g_2)=\Big(\begin{smallmatrix}
1 & 0 & 0\\
0 & -1 & 0\\
0 & 0 & -1
\end{smallmatrix}\Big)$}, $\mathcal{D}(g_1) {\bf M}={\bf M}$, whereas the group elements $g_2$ and $g_1g_2$ change the sign of the $y$ and $z$ component of ${\bf M}$, $\mathcal{D}(g_2) {\bf M}=\mathcal{D}(g_1g_2){\bf M}=(M_x,-M_y,-M_z) \equiv (M_x,M_y,M_z)$, both the $M_y$ and $M_z$ must be zero. Therefore, the net magnetic moment of RuF$_4$ must point along the $x$-axis, in accordance with our DFT results. 

The transition from the altermagnetic to the weak ferromagnetic phase leaves its fingerprints on \textit{the spin texture}. Spin texture is a vector field in the reciprocal space defined through the expectation value of the spin operator ${\bf s} = (s_x, s_y, s_z)$ in a given Bloch state $|\psi_{k_x,k_y}^n\rangle$,
\begin{equation}
     \langle s_i \rangle^n_{k_x,k_y} = \langle \psi_{k_x,k_y}^n|s_i|\psi_{k_x,k_y}^n\rangle, \quad i=x,y,z.
\end{equation}
The spin texture, like the net magnetic moment, complies with the symmetry of the system. We provide detailed derivation in Appendix~\ref{SOCgeneral} and arrive here straight at the conclusion by expressing the action of the group elements on the spin texture,
\begin{eqnarray}\label{eq:spin-textures}
    g_1:&&  \langle s_i \rangle^n_{k_x,k_y}=\langle s_i \rangle^n_{-k_x,-k_y}, \nonumber\\
    g_2:&&  \langle s_i \rangle^n_{k_x,k_y}=(-1)^{1-\delta_{i,x}}\langle s_i \rangle^n_{k_x,-k_y}, \nonumber\\
    g_1g_2:&& \langle s_i \rangle^n_{k_x,k_y}=(-1)^{1-\delta_{i,x}}\langle s_i \rangle^n_{-k_x,k_y},
\end{eqnarray}
where $\delta_{i,x}$ is the Kronecker delta, equal to 1 if $i=x$ and 0 otherwise. 

The Eq.~\ref{eq:spin-textures} shows that the spin texture in the I quadrant of the BZ $(k_x>0,k_y>0)$ is related to the spin texture in the II, III, and IV through the action of the group elements $g_1g_2$, $g_1$, and $g_2$, respectively. 
We can illustrate these relations with a concrete example. In Fig.~\ref{fig:canting}c we show the spin texture of the valence bands V$_{1/2}$ at the constant energy surface $-0.15 \, {\rm eV}$ below the Fermi level.  
Now, it is clear that the spin expectation values $\langle s_{y/z} \rangle_{k_x,k_y}$ from the II and IV quadrants cancel out the spin expectation values in the I and III quadrants (actually $\langle s_z \rangle$ is zero everywhere, which is not shown in the plot). The cancellation of the spin expectation values from different quadrants of the BZ leads to the vanishing of $M_y$ and $M_z$ in the real space. Indeed, each component of the net magnetic moment can be expressed as an integral of the corresponding spin expectation values over the BZ, $M_i = \sum_n^{{\rm occ}} \int_{BZ} \langle s_i \rangle^n_{k_x,k_y} {\rm d}^2{\bf k}$, where the "occ" denotes that the sum runs over the occupied states. On the other hand, $\langle s_{x} \rangle_{k_x,k_y} > 0$ in all the four quadrants,  yielding non-zero $M_x$.

In the following, we question the microscopic mechanism responsible for canting the magnetic moments. In the work of Wang \textit{et al}\cite{Wang2022Aug} it is shown that the Dzyaloshinskii-Moriya interaction (DMI) acting between the Ru moments is mainly responsible for their canting, with some contribution from the single-ion anisotropy (SIA). The DMI between the opposite-spin Ru atoms is allowed because the Ru-Ru bond does not display inversion\cite{Moriya1960Oct}, as we mentioned in the Subsection~\ref{ssec_symmetry_analysis}. Given that SOC is the common origin of both the DMI and SIA, the magnitude of the induced net magnetic moment must depend on the SOC strength. To shed light on this relation, we scale the SOC strength in DFT calculations by using the modified SOC constant $\widetilde{\lambda} = \xi \lambda$, where the dimensionless parameter $\xi$ ranges from 0 (no SOC) to 1 (realistic SOC). Then, we constrain the magnetic moments in the directions that are canted by angle $\alpha$ from the $y$-axis. For a set of directions $\alpha$ the equilibrium canting angle $\alpha_{\xi}$ is found by minimizing the auxiliary function $E_{\xi}(\alpha) = c_0 + c_2 (\alpha-\alpha_{\xi})^2$, where $E_{\xi}(\alpha)$ is the total energy of the system for a particular canting angle $\alpha$ and SOC strength $\widetilde{\lambda} = \xi \lambda$. Finally, for each $\xi$ we perform DFT calculations for magnetic moments canted by $\alpha_{\xi}$ and we calculate the net magnetic moment. The plot in Fig.~\ref{fig:canting}c shows that the net magnetic moment is linear in SOC strength. 

The linearity of the dependence $M(\tilde{\lambda})$ can be explained by turning to the microscopic expression for the Dzyaloshinskii-Moriya interaction. First, we noticed that the magnitude of the Ru cluster's magnetic moment ${\bf m} = {\bf m}({\rm Ru}) + 2 {\bf m}({\rm F})$  is almost unaffected by SOC, as the magnitude of the magnetic moment on Ru atom changes only slightly from $1.464 \, \mu_{\rm B}$ for $\xi = 0$ to $1.453 \, \mu_{\rm B}$ for $\xi = 1$ while $|{\bf m}({\rm F})|$ doesn't change at all, staying $0.214 \, \mu_{\rm B}$. Second, the magnitude of the net magnetic moment ${\bf M} = {\bf m}_A + {\bf m}_B$ and the canting angle $\alpha_{\xi}$ are related by $M = 2 m \sin \alpha_{\xi}$, where $m = |{\bf m}_{A/B}|$. Now, if we assume that canted magnetic moments stay in the $xy$-plane, the DMI energy can be expressed as
\begin{equation}
    E_{\rm DMI} = {\bf D} \cdot ({\bf m}_A \times {\bf m}_B) = D \cos \theta_D \sin (2 \alpha_{\xi}),
\end{equation}
where $\theta_D = \measuredangle ({\bf D}, z)$. As pointed out by Moriya\cite{Moriya1960Oct} the magnitude of the Dzyaloshinskii vector is linear in SOC constant, $D \sim \lambda$ (here $D \sim \tilde{\lambda}$ as we are using modified SOC constant), whereas its direction depends on the symmetry of the structure and on the atomic positioning. Therefore, the angle $\theta_{D}$ does not depend on $\tilde{\lambda}$ and the DMI energy can be expressed as $E_{\rm DMI} \sim \tilde{\lambda} \sin (2 \alpha_{\xi})$. As long as the canting angle is small the approximate relation $E_{\rm DMI} \sim 2 \tilde{\lambda} \alpha_{\xi}$  holds. Noticing that the DMI energy is quadratic in Dzyaloshinskii vector magnitude\footnote{For a quantum spin-1 system described by Hamiltonian $\mathcal{H} = J{\bf S}_1 \cdot {\bf S}_2 + {\bf D} \cdot ({\bf S}_1 \times {\bf S}_2)$, ${\bf D}=D{\bf e}_z$,  the DMI correction of the eigenvalues is equal to $\pm\sqrt{J^2 + D^2} \sim \pm J\big( 1+\frac{D^2}{2J^2}\big)$ and $\frac{J}2(1\pm3\sqrt{1 + \frac{8D^2}{9J^2}}) \sim \frac{J}2(1\pm3(1+\frac{4D^2}{9J^2}))$, where it is assumed that the DMI is much smaller than the Heisenberg exchange. Therefore, the DMI correction to the energy is $E_{\rm DMI} \sim D^2/J$.}, $E_{\rm DMI} \sim D^2$ , the last relation shows that the canting angle is linear in the modified SOC constant, $\alpha_{\xi} \sim \tilde{\lambda}$. Finally, for small canting angles, the linearity of the $M(\tilde{\lambda})$ dependence stems from the relation $M \sim 2m \alpha_{\xi} \sim m \tilde{\lambda}$.

\subsection{SOC-induced spin splitting of bands}
\label{ssec.band_splitting}
\begin{figure}[h!]
\centering
\includegraphics[width=0.9\linewidth]{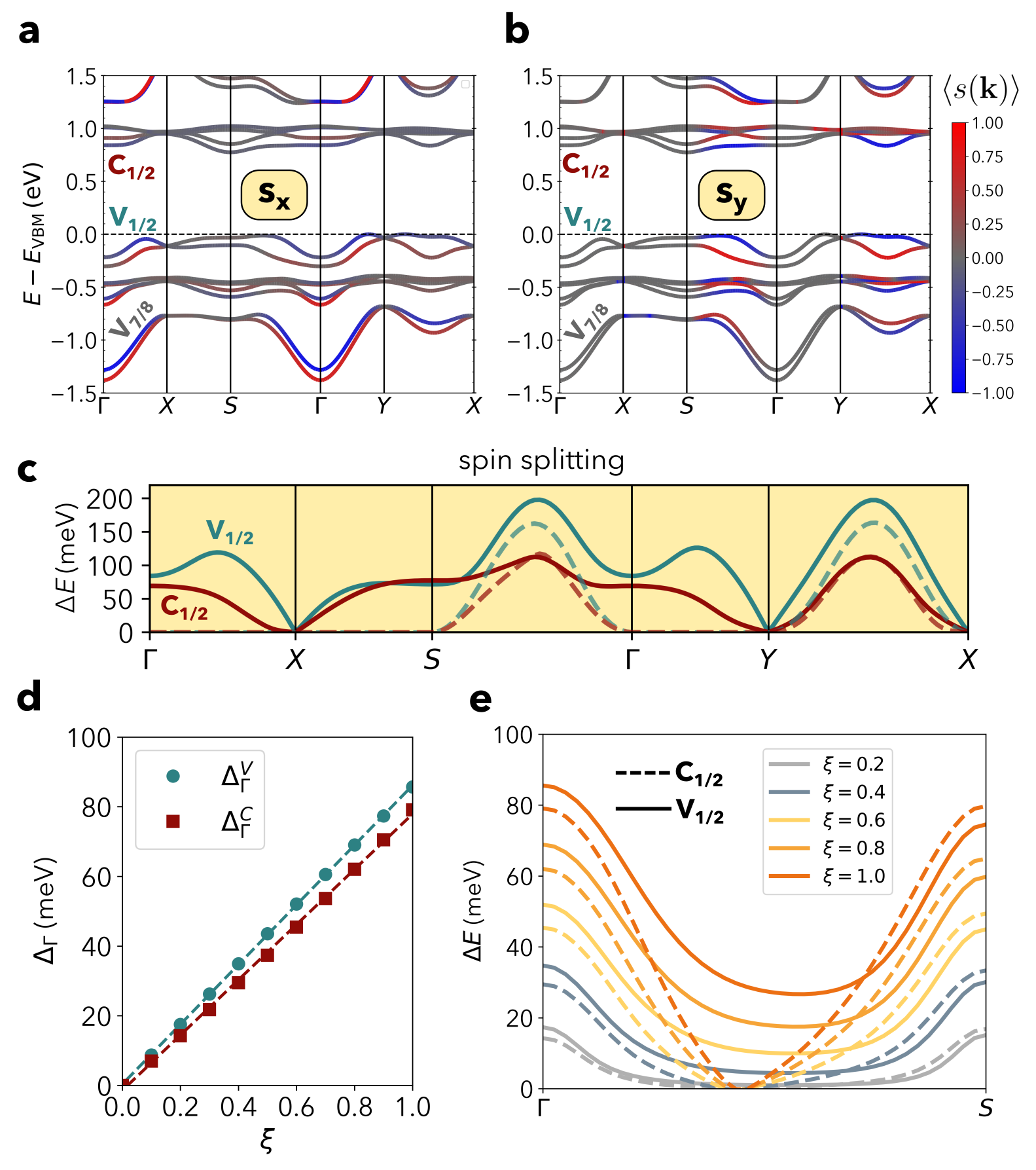}
\caption{\label{fig:bands-AM-SOC} 
Non-collinear relativistic DFT calculations of the band structure for RuF$_4$ along high symmetry lines.
(a)~spin expectation values $\langle s_x({\bf k})\rangle$ and (b)~$\langle s_y({\bf k})\rangle$ plotted as a band color. 
(c)~spin splitting of the top valence V$_{1/2}$ and bottom conduction bands C$_{1/2}$ is along the same $k$-path given, with the AM-induced spin splitting plotted for comparison with dashed lines.
(d)~spin splitting of V$_{1/2}$ and C$_{1/2}$ at the $\Gamma$ point calculated for the different SOC strength  $\xi$.
(e)~the difference between spin splitting energies with and without SOC of V$_{1/2}$ and C$_{1/2}$ bands along the $\Gamma$S path for different SOC strengths $\xi$.
}
\end{figure}

Spin-orbit coupling, besides inducing net magnetic moment in the 2D altermagnet RuF$_4$, is responsible for significant spin splitting of the electronic bands. To shed light on this SOC effect, in Fig.~\ref{fig:bands-AM-SOC}(a)-(b) we plot the band structure calculated with SOC along the same $k$-path and in the same energy window as in the Fig.~\ref{fig:bands-AM}. In addition, we project the spin expectation values $\langle {s_x}\rangle_{k_x,k_y}$ and $\langle {s_y}\rangle_{k_x,k_y}$ over each band to reveal how their spin polarization evolves along different $k$-directions. 

Due to SOC, the bands V$_{1/2}$ and C$_{1/2}$, which are degenerate in the AM phase along some special directions, are now split in almost the entire BZ, displaying highly nonuniform splitting (Fig.~\ref{fig:bands-AM-SOC}c). Intriguingly, the non-relativistic (AM-induced) and relativistic (SOC-induced) spin splitting reach their respective maxima in different regions of the BZ: while SOC-induced splitting is maximal at $\Gamma$ and S points, where the bands were previously degenerate in the AM phase, the AM-induced splitting in the middle of the $\Gamma$S and XY lines is barely affected by SOC.  

Apart from the splitting energies, the polarization of bands displays non-uniform $k$-dependence. For example, the V$_{1/2}$ bands are polarized in the $x$-direction along the $\Gamma$X and $\Gamma$Y lines, with $\langle s_x \rangle < 0$ for V$_1$ and $\langle s_x \rangle > 0$ for V$_2$ (Fig.~\ref{fig:bands-AM-SOC}a-b). The polarization in the $x$-direction along these $k$-lines is the most evident for the V$_{7/8}$  bands. On the other hand, both $\langle s_x \rangle$ and $\langle s_y \rangle$ are nonzero and vary substantially along the $\Gamma$S and XY lines, meaning that V$_{1/2}$ have mixed polarization along these directions. Also, there are $k$-directions along which the bands are almost unpolarized, such as the XS line where $\langle s_x \rangle \approx \langle s_y \rangle \approx 0$. 

The point that deserves special attention is the lifting of spin degeneracy along the X$\Gamma$Y path. What is particularly striking is a large spin splitting at $\Gamma$, which is a time-reversal invariant momenta (TRIM) point, of $\Delta_{\Gamma}^{V_{1/2}}=86 \, {\rm meV}$ and $\Delta_{\Gamma}^{C_{1/2}}=79 \, {\rm meV}$ for the V$_{1/2}$ and C$_{1/2}$ bands, respectively. Moreover, $\Delta_{\Gamma}$ scales linearly with the SOC strength, as shown in Fig.~\ref{fig:bands-AM-SOC}d. At first glance, this linear splitting seems to occur due to a nonzero net magnetic moment pointing in the $x$-direction, but our additional DFT calculations reveal that even when the Ru magnetic moments are strictly constrained in the $y$-direction (so that ${\bf M} = 0$), the spin splitting of bands along X$\Gamma$Y persists. 

To shed light on this, let us consider a non-relativistic Hamiltonian of the AM phase $\mathcal{H}^{\rm nrel}_{\rm AM}$ and include SOC as a perturbation. The complete derivation is exposed in Appendix \ref{SOCperturbation} but here we discuss the most important conclusions. Assuming that two Bloch wavefunctions corresponding to the same ${\bf k} = (k_x, k_y)$ point, $\ket{\Psi_1 }_{k_x,k_y}=\ket{\psi_1}\otimes\ket{\uparrow}$ and $\ket{\Psi_2}_{k_x,k_y}=\ket{\psi_2}\otimes\ket{\downarrow}$, are with the same energy $E$, the perturbation $\mathcal{H}_{\rm SOC}= ({\bm \nabla} V \times {\bf p})\cdot {\bm \sigma} = \lambda {\bf L}\cdot {\bm \sigma}$ in the $\{\ket{\Psi_1}_{k_x,k_y}, \ket{\Psi_2}_{k_x,k_y}\}$ basis has the matrix elements of the form
\begin{equation}
\label{eq.SOC-matrix}
 \langle \mathcal{H}_{\rm SOC} \rangle =  \lambda \begin{pmatrix}
        \ell_y & {\rm i}(\ell_x - \ell_z) \\
   -{\rm i}(\ell_x -\ell_z)&  \ell_y
    \end{pmatrix},
\end{equation}
where $\ell_i = \ell_i(k_x, k_y)$, $i=x,y,z$, are real numbers that can be connected to nonzero matrix elements of the pseudovector operator ${\bf L}$, see Appendix~\ref{SOCperturbation}. From the Eq.~\ref{eq.SOC-matrix}, the SOC contribution to the level $E$ is $\lambda\Big(\ell_y\pm (\ell_x - \ell_z)\Big)$, meaning that SOC lifts the degeneracy of $\ket{\Psi_1}_{k_x,k_y}$ and $\ket{\Psi_2}_{k_x,k_y}$ inducing the spin splitting of $\Delta_{\bf k} = 2\lambda (\ell_x - \ell_z)$. The derivation exposed in Appendix~\ref{SOCperturbation} applies to any two degenerate Bloch wavefunctions as it only exploits the action of the group element $g_2$ and the assumption that ${\bf N}$ points in the $y$-direction. Therefore, the conclusions of Appendix~\ref{SOCperturbation} apply to any ${\bf k}$ point, showing that the linear-in-SOC spin splitting is a cooperative effect of breaking the time-reversal symmetry due to the magnetic order and of the spin-orbit coupling. 
 
The spin degeneracy along the $\Gamma$S and XY lines is lifted already in the AM phase, but the inclusion of SOC changes this spin splitting drastically. To show how the SOC-induced splitting gradually develops over the AM-induced one, we plot the difference of the SOC-induced and AM-induced splitting energies for V$_{1/2}$ and C$_{1/2}$ bands along the $\Gamma$S path for different SOC strengths $\tilde{\lambda} = \xi \lambda$ ( Fig.~\ref{fig:bands-AM-SOC}e).
As $\xi$ varies from 0 to 1 the SOC-induced splitting increases, but the fraction of the  $\Gamma$S path where it dominates over the AM-induced splitting never surpasses $40\%$. Therefore, the AM-induced splitting is still dominant along the $\Gamma$S line despite the sizable SOC-induced spin splitting. 

\section{Discussion and conclusion}\label{sec.conclusion}
\noindent

In this work we have shown that in the non-relativistic limit, the monolayer RuF$_4$ represents a two-dimensional $d$-wave altermagnet. Using the type-I magnetic space group, we have revealed the symmetry properties of the non-relativistic spin splitting in the altermagnetic phase. By applying the theory of invariants, we have derived a simple model Hamiltonian that captures the spin-momentum locking of the top valence and the bottom conduction bands, both of which can be activated for spin transport by doping. In the relativistic limit, we have shown that spin-orbit coupling induces weak ferromagnetism and significantly changes the spin splitting and the spin texture of bands. Using the perturbation theory, we explained the appearance of a sizable SOC-induced spin splitting at the time-reversal invariant $\Gamma$ point, which is reminiscent of Zeeman splitting but occurs even in the limit of vanishing net magnetic moment. We have shown that the magnetic moment induced due to canting of Ru spins is linear in SOC strength. This implies that in altermagnetic compounds containing heavy atoms, which exhibit very strong SOC, the altermagnetic phase could be significantly altered
by weak ferromagnetism. 

Finally, it's worth noting that Autieri et al.\cite{Autieri2023Dec} recently highlighted that the emergence of weak ferromagnetism in metallic altermagnets is a property unparalleled in conventional antiferromagnets. As such, the appearance of weak magnetization in metallic compounds can be used to demarcate the conventional antiferromagnets from altermagnets. Yet, as we have shown in this work, RuF$_4$ is a semiconductor and we argue that the appearance of SOC-induced weak ferromagnetism may be a general property of all the altermagnetic compounds, not just metals. This becomes especially encouraging once one realizes that the small component of the magnetic moment, induced by SOC and perpendicular to the N\'{e}el vector, could be employed as a knob to control the direction of dominant magnetic arrangement in altermagnets. Indeed, the N\'{e}el vector in weak ferromagnets can be dragged by a small external magnetic field using the experimental technique exposed in the work of Dmitrienko \textit{et al.}\cite{Dmitrienko2014Mar}. In this way one could hopefully control the direction of spin-polarized currents in altermagnets. Therefore, we strongly believe that the relation between weak ferromagnetism and altermagnetism warrants greater attention, and would stimulate interest to delve into this intriguing topic.

\section{Methods}
DFT calculations were performed using the {\sc VASP} code \cite{Kresse1996Oct}. The effects of electronic exchange and correlation were described at the Generalized Gradient Approximation (GGA) level using the Perdew–Burke–Ernzerhof (PBE) functional \cite{Perdew1996Oct}. The Kohn-Sham wavefunctions were expanded on a plane wave basis set with a cutoff of $500 \, {\rm eV}$. We used the pseudopotentials with Ru 4\textit{p},5\textit{s},4\textit{d} and F 2$s$,2$p$ states as valence states. The lattice constants $a_1 = 5.131 \, {\rm \AA}$ and $a_2 = 5.477 \, {\rm \AA}$ were obtained from spin-polarized non-relativistic DFT calculations using the Birch-Murnaghan equation of state and assuming the antiparallel arrangement of Ru magnetic moments. In the out-of-plane direction, we used the lattice constant of $a_3 = 20 \, {\rm \AA}$ to ensure sufficient separation between the periodic replicas. The atomic positions were relaxed until all the forces' components dropped below $0.002 \, {\rm eV/\AA^2}$. The Brillouin zone was sampled with $10 \times 9 \times 1$ ${\bf k}$ points mesh during the relaxation and with a finer $15 \times 14 \times 1$ mesh during the self-consistent field calculations. To find the canting angle of the Ru magnetic moment we performed non-collinear DFT calculations with SOC and for a set of directions we constrained the magnetic moments using the penalty functional~\cite{Ma2015Feb}. For producing the bandstructure and spin texture plots we used the Pyprocar package\cite{Herath2020Jun}. For setting up DFT calculations and for structural visualization we used the Atomic Simulation Environment \cite{Larsen2017Jun}, {\sc XCrySden}\cite{Kokalj1999Jun}, and {\sc Vesta}\cite{Momma2011Dec} packages.  

\acknowledgments
We acknowledge support by the Italian Ministry for Research and Education through the Nanoscience Foundries and Fine Analysis (NFFA-Trieste, Italy) project and through the PRIN-2017 project "TWEET: Towards ferroelectricity in two dimensions" (IT-MIUR grant No. 2017YCTB59). M.M, M.O., and S.S.~acknowledge the financial support provided by the Ministry of Education, Science, and Technological Development of the Republic of Serbia. This project has received funding from the European Union's Horizon 2020 Research and Innovation Programme under the Programme SASPRO 2 COFUND Marie Sklodowska-Curie grant agreement No. 945478. 
M.G.~acknowledges financial support provided by Slovak Research and Development Agency provided under Contract No. APVV-SK-CZ-RD-21-0114 and by the Ministry of Education, Science, Research and Sport of the Slovak Republic provided under Grant No. VEGA 1/0695/23 and Slovak Academy of Sciences project IMPULZ IM-2021-42 and project FLAG ERA JTC 2021 2DSOTECH. The authors gratefully acknowledge the Gauss Centre for Supercomputing e.V. for providing computational resources on the GCS Supercomputer SuperMUC-NG at Leibniz Supercomputing Centre and the Italian supercomputing centre CINECA for providing computational resources through the ISCRA initiative, in particular projects ISCRA-C HP10C6WZ1O and ISCRA-B HP10BA00W3. 

\section*{References}
\bibliography{references}

\appendix
\section{General constraints on spin expectation values in the relativistic case}\label{SOCgeneral}

Let $\ket{\psi_{k_x,k_y}^n}$ represents an eigenstate of the altermagnetic Hamiltonian in a relativistic phase $\mathcal{H}_{AM}^{\rm rel}$
\begin{equation}\label{Bloch}
\mathcal{H}_{AM}^{\rm rel}\ket{\psi_{k_x,k_y}^n}=E_n(k_x,k_y)\ket{\psi_{k_x,k_y}^n},
\end{equation}
with group symmetry ${\bf G}=p2_1/b11$ of both the relativistic and/or altermagnetic phase, since the magnetic space group is equal to the crystallographic group, see Section~\ref{ssec.nonrel_description}.
The commutation relation $[d(g),\mathcal{H}_{AM}^{\rm rel}]=0$ holds for any element $g \in {\bf G}$, where $d$ is the representation of the group acting in the Hilbert space of the Hamiltonian $\mathcal{H}_{AM}^{\rm rel}$.
For a given state $\ket{\psi_{k_x,k_y}^n}$, the spin expectation values are defined as
\begin{equation}
     \langle s_i \rangle^n_{k_x,k_y}=\langle \psi_{k_x,k_y}^n|s_i|\psi_{k_x,k_y}^n\rangle, \quad i=x,y,z.
\end{equation}
Using the action of the group element $g\in{\bf G}$ in the Hilbert space in which the Hamiltonian $\mathcal{H}$ acts, the previous equation can be transformed to
\begin{equation}
     \langle s_i \rangle^n_{k_x,k_y}=\langle \psi_{k_x,k_y}^n|d^{\dag}(g)d(g)s_id^{\dag}(g)d(g)|\psi_{k_x,k_y}^n\rangle
\end{equation}
By analyzing the action of nontrivial group elements $g_1=\mathcal{I}$, $g_2=(U_x|t_{\frac{1}{2}})$
and $g_1g_2$, one obtains general constraints on the spin expectation values
\begin{eqnarray}\label{generalSPINEXP}
    g_1:&&  \langle s_i \rangle^n_{k_x,k_y}=\langle s_i \rangle^n_{-k_x,-k_y}, \nonumber\\
    g_2:&&  \langle s_i \rangle^n_{k_x,k_y}=(-1)^{1-\delta_{i,x}}\langle s_i \rangle^n_{k_x,-k_y}, \nonumber\\
    g_1g_2:&& \langle s_i \rangle^n_{k_x,k_y}=(-1)^{1-\delta_{i,x}}\langle s_i \rangle^n_{-k_x,k_y},
\end{eqnarray}
where $\delta_{i,x}$ is the Kronecker delta,
equal to 1 if $i=x$ and 0 otherwise. To obtain the previous relations, we have used the following identities
\begin{eqnarray}
    g_1:&& d(g_1)s_id^{\dag}(g_1)=s_i,\phantom{xxxxxxxxxx}d(g_1)|\psi_{k_x,k_y}^n\rangle=|\psi_{-k_x,-k_y}^n\rangle,\nonumber\\
    g_2:&& d(g_2)s_id^{\dag}(g_2)=(-1)^{1-\delta_{i,x}}s_i,\phantom{xxx}d(g_2)|\psi_{k_x,k_y}^n\rangle=|\psi_{k_x,-k_y}^n\rangle\nonumber\\
     g_1g_2:&& d(g_1g_2)s_id^{\dag}(g_1g_2)=(-1)^{1-\delta_{i,x}}s_i,d(g_1g_2)|\psi_{k_x,k_y}^n\rangle=|\psi_{-k_x,k_y}^n\rangle.
\end{eqnarray}
In short, the set of equations~\ref{generalSPINEXP} show that in a system with $p2_1/b11$ symmetry the spin texture $\langle s_x \rangle$ can accumulate, yielding a finite net magnetic moment in the $x$-direction, whereas the $\langle s_y \rangle$ and $\langle s_z \rangle$ from different quadrants of the BZ cancel out, as explained in the main text and illustrated in Fig.~\ref{fig:canting}b. 

Furthermore, the equations~\ref{generalSPINEXP} in the special cases of the $\Gamma$ point and  along the $\Gamma$X and $\Gamma$Y lines, suggest that spin expectation values are nonzero in the $x$-direction solely since for $k_y=0$ ($\Gamma$X line) we have
\begin{eqnarray}
     \langle s_x \rangle^n_{k_x,0}&=& \langle s_x \rangle^n_{k_x,0},\nonumber\\
     \langle s_y \rangle^n_{k_x,0}&=& -\langle s_y \rangle^n_{k_x,0}=0,\nonumber\\
     \langle s_z \rangle^n_{k_x,0}&=& -\langle s_z \rangle^n_{k_x,0}=0,
\end{eqnarray}
while along the $\Gamma$Y line ($k_y=0$) one gets
\begin{eqnarray}
     \langle s_x \rangle^n_{0,k_y}&=& \langle s_x \rangle^n_{0,k_y},\nonumber\\
     \langle s_y \rangle^n_{0,k_y}&=& -\langle s_y \rangle^n_{0,k_y}=0,\nonumber\\
     \langle s_z \rangle^n_{0,k_y}&=& -\langle s_z \rangle^n_{0,k_y}=0.
\end{eqnarray}
This conclusion is in agreement with the band structure plotted in Fig.~\ref{fig:bands-AM-SOC} which shows that along the X$\Gamma$Y  path the $x$-component of spin is nonzero (blue and red bands Fig.~\ref{fig:bands-AM-SOC}a) while the $y$-component equals to zero (gray bands Fig.~\ref{fig:bands-AM-SOC}b).

Also, it should be mentioned that the conclusions agree with the ones given in Section~\ref{SOCperturbation},
where we have shown that the SOC is solely responsible for the observed splitting along the high-symmetry lines $\Gamma$X and $\Gamma$Y, as well as in the $\Gamma$ point.

\section{Perturbation theory of SOC-induced removal of spin degeneracy at arbitrary $k$ point}\label{SOCperturbation}

The spin splitting at the $\Gamma$ point and along the $\Gamma$X and $\Gamma$Y lines can be explained using the perturbation theory, where the SOC Hamiltonian is introduced as a perturbation. To do this, we apply a first-order degenerate perturbation theory, valid for bands not only along the X$\Gamma$Y path, but also at other $k$ points of the Brillouin zone which host the degenerate bands in the altermagnetic phase.

We describe the two degenerate states as $\ket{\Psi_1}=\ket{\psi_1}\otimes\ket{\uparrow}$ and $\ket{\Psi_2}=\ket{\psi_2}\otimes\ket{\downarrow}$, where $\ket{\psi_{1/2}}$ corresponds to the orbital part of the wave function,
while  $\ket{\uparrow / \downarrow}$ are eigenstates of the Pauli $\sigma_y$ operator for the eigenvalues 1 and $-1$, respectively. We chose $\sigma_y$ because we assumed that the spin quantization axis is along the $y$-direction. Now, if the $\mathcal{H}^{\rm nrel}_{\rm AM}$ is the non-relativistic one-electron Hamiltonian (AM emphasizes here that this is the non-relativistic Hamiltonian of the altermagnetic phase), then the following relation holds
\begin{equation}
\mathcal{H}^{\rm nrel}_{\rm AM}\ket{\Psi_{1/2}}=E\ket{\Psi_{1/2}}.
\end{equation}
In the orbital space the action of the group element $g_2$ is through the representation $d_O(g_2)$, while in the magnetic space acts as a representation $d_M$, $d_M(g_2)=\big(\begin{smallmatrix}
    0 & 1 \\
    1 & 0
\end{smallmatrix}\big)$, which flips the magnetic moments $d_M(g_2)\ket{\uparrow/\downarrow}=\ket{\downarrow/\uparrow}$. Using the commutation relation between the Hamiltonian and the elements of the group, $[d_O(g_2)\otimes d_M(g_2),\mathcal{H}^{\rm nrel}_{\rm AM}]=0$, one can say that
\begin{eqnarray}
    \Big(d_O(g_2)\otimes d_M(g_2)\Big) \mathcal{H}^{\rm nrel}_{\rm AM}\ket{\psi_{1}}\otimes\ket{\uparrow}&=&\Big(d_O(g_2)\otimes d_M(g_2)\Big)E \ket{\psi_{1}}\otimes\ket{\uparrow}\nonumber\\
   \mathcal{H}^{\rm nrel}_{\rm AM} \Big(d_O(g_2)\ket{\psi_{1}}\Big)\otimes\ket{\downarrow}&=&E\Big(d_O(g_2)\ket{\psi_{1}}\Big)\otimes\ket{\downarrow}.
\end{eqnarray}
The last relation is equal to $\mathcal{H}^{\rm nrel}_{\rm AM}\ket{\psi_{2}}\otimes\ket{\downarrow}=E\ket{\psi_{2}}\otimes\ket{\downarrow}$, which leads us to conclude that the relation $d_O(g_2)\ket{\psi_1}=\ket{\psi_2}$ must hold. Furthermore, using the fact that $d_O(g_2)d_O(g_2)=d_O(g_2^2) =I$, where $I$ is the identity matrix, one can conclude that
$d_O(g_2^2)\ket{\psi_1}=\ket{\psi_1}$, which suggests that the relation $d_O(g_2)\ket{\psi_2}=\ket{\psi_1}$ holds. Therefore, the group element $g_2$ interchanges the orbital parts of two degenerate states. This will be useful for calculating the matrix elements of the SOC operator in the $\{|\Psi_1\rangle, |\Psi_2\rangle\}$ basis.  

To account for the SOC effects we use the first-order perturbation theory and calculate the allowed matrix elements of the operator $H_{\rm SOC}=({\bm \nabla} V \times {\bf p})\cdot {\bm \sigma}$, where ${\bm \nabla} V$ represents the gradient of the crystal potential, ${\bf p}$ the momentum operator, and ${\bm \sigma}$ the Pauli operator. Now, the fact that $({\bm \nabla} V \times {\bf p})$ transforms as a pseudovector allows us to rewrite the SOC Hamiltonian as $H_{\rm SOC}=\lambda {\bf L}\cdot {\bm \sigma}$, where $\lambda$ describes the SOC strength. Note that ${\bf L}$ \textit{is not} the angular momentum but it transforms like one. Now, the matrix elements of $H_{\rm SOC}$ in the $\{|\Psi_1\rangle, |\Psi_2\rangle\}$ basis are 
\begin{equation}
 \langle H_{\rm SOC} \rangle =  \lambda \begin{pmatrix}
        \langle \psi_1|L_y|\psi_1 \rangle & {\rm i}\langle \psi_1|L_x|\psi_2 \rangle- \langle \psi_1|L_z|\psi_2 \rangle \\
    -{\rm i}\langle \psi_2|L_x|\psi_1 \rangle-\langle \psi_2|L_z|\psi_1 \rangle &  -\langle \psi_2|L_y|\psi_2 \rangle 
    \end{pmatrix}.
\end{equation}
Here we have used that the Pauli operators in a given spin basis $\{\ket{\uparrow}=\frac{1}{\sqrt{2}}\big(-{\rm i}\ket{+}+\ket{-}\big),\ket{\downarrow} = \frac{1}{\sqrt{2}}\big({\rm i}\ket{+}+\ket{-}\big)\}$ ($\ket{\pm}$ are eigenvectors of the $\sigma_z$ operator, $\sigma_z\ket{\pm}=\pm\ket{\pm}$) are equal to $\sigma_x=\big(\begin{smallmatrix}
          0 & {\rm i}\\
   {-\rm i} & 0
\end{smallmatrix}\big)$, $\sigma_y=\big(\begin{smallmatrix}
          1 & 0\\
           0 & -1
\end{smallmatrix}\big)$, $\sigma_z=\big(\begin{smallmatrix}
          0 &  -1\\
          -1 & 0
\end{smallmatrix}\big)$. The symmetry argument in this case gives us
\begin{eqnarray}
    \langle \psi_1|L_y|\psi_1 \rangle&=& \langle \psi_1|d_O^{\dag}(g_2)d_O(g_2)L_yd_O^{\dag}(g_2)d_O(g_2)|\psi_1 \rangle=-\langle \psi_2|L_y|\psi_2 \rangle=\ell_y,
\end{eqnarray}
where we have used that $d_O(g_2)L_yd_O^{\dag}(g_2)=-L_y$ and that $\ell_y$ is the real number. Also, we calculate the matrix elements $\langle \psi_1|L_x|\psi_2 \rangle$ and $\langle \psi_1|L_z|\psi_2 \rangle$,
\begin{eqnarray}\label{LxLzconstraint}
    \ell_x^{12}=\langle \psi_1|L_x|\psi_2 \rangle&=& \langle \psi_1|d_O^{\dag}(g_2)d_O(g_2)L_xd_O^{\dag}(g_2)d_O(g_2)|\psi_2 \rangle=\langle \psi_2|L_x|\psi_1 \rangle=(  \ell_x^{12})^*,\nonumber\\
       \ell_z^{12}=\langle \psi_1|L_z|\psi_2 \rangle&=& \langle \psi_1|d_O^{\dag}(g_2)d_O(g_2)L_zd_O^{\dag}(g_2)d_O(g_2)|\psi_2 \rangle=-\langle \psi_2|L_z|\psi_1 \rangle=- (\ell_z^{12})^*,\nonumber\\
\end{eqnarray}
and use the relations $d_O(g_2)L_xd_O^{\dag}(g_2)=L_x$ and
$d_O(g_2)L_zd_O^{\dag}(g_2)=-L_z$.

The conditions obtained in~\eqref{LxLzconstraint} imply that $\ell_x^{12}=\ell_x\in \mathcal{R}$ and $\ell_z^{12}={\rm i}\ell_z$,
where $\ell_z\in \mathcal{R}$, allowing us to simplify 
$\langle H_{\rm SOC} \rangle $ to 
\begin{equation}\label{SOCfinal}
 \langle H_{\rm SOC} \rangle =  \lambda \begin{pmatrix}
        \ell_y & {\rm i}(\ell_x - \ell_z) \\
   -{\rm i}(\ell_x -\ell_z)&  \ell_y
    \end{pmatrix}.
\end{equation}
Using the final form of $\langle H_{\rm SOC}\rangle$ given by Eq.~\eqref{SOCfinal}, the SOC-induced energy contribution to the degenerate level in the altermagnetic case is equal $\lambda\Big(\ell_y\pm (\ell_x - \ell_z)\Big)$. The spin splitting due to the  SOC is equal to $2\lambda (\ell_x - \ell_z)$. Furthermore, by calculating the eigenvectors of $E_0 I_2+ \langle H_{\rm SOC} \rangle$, where $E_0$ represents the energy of the double degenerate bands, while $I_2$ is the $2\times2$ identity matrix in the eigenbasis $\{\ket{\uparrow},\ket{\downarrow}\}$ of the altermagnetic phase, we find that the spin expectation values have the nonzero component in the $x$-direction only, consistent with the general conclusion from Appendix~\ref{SOCgeneral}. However, the first-order perturbation theory predicts $\langle s_x \rangle=\pm 1$ (in the units of $\hbar$/2) at the $\Gamma$ point, which is not consistent with the data obtained from first-principles calculations, where $\langle s_x \rangle < 1$. This implies that the higher-order perturbation theory needs to be applied to account for the interaction with the bands higher in energy, responsible for the reduction of the intensity of the spin expectation value.

\end{document}